# Inverse Raman Scattering in Silicon

Daniel R. Solli, Prakash Koonath and Bahram Jalali
Department of Electrical Engineering, University of California, Los Angeles
Los Angeles, CA 90095-1594

**Abstract:**

Stimulated Raman scattering is a well-known nonlinear process that can be harnessed to produce optical gain in a wide variety of media. This effect has been used to produce the first silicon-based lasers and high-gain amplifiers. Interestingly, the Raman effect can also produce intensity-dependent nonlinear *loss* through a corollary process known as inverse Raman scattering (IRS). Here, we demonstrate IRS in silicon—a process that is substantially modified by the presence of optically-generated free carriers—achieving attenuation levels >15 dB with a pump intensity of 4 GW/cm$^2$. Ironically, we find that free-carrier absorption, the detrimental effect that suppresses other nonlinear effects in silicon, actually facilitates IRS by delaying the onset of contamination from coherent anti-Stokes Raman scattering. The carriers allow significant IRS attenuation over a wide intensity range. Silicon-based IRS could be used to produce chip-scale wavelength-division multiplexers, optical signal inverters, and fast optical switches.

Raman scattering is an inelastic process that is extremely useful for spectroscopic analysis, and can also be used to produce optical gain in a wide variety of transparent media. In its spontaneous form, intense pump light is injected into an optical medium, and a minute amount is scattered to redshifted (Stokes) wavelengths by material vibrations. When the process is stimulated by radiation at the Stokes wavelength, it becomes much stronger, as the Stokes wave experiences amplification[1]. As such, stimulated Raman scattering (SRS) has been instrumental in producing silicon-based lasers and amplifiers[2,3,4,5]. The Raman effect also couples the pump and Stokes waves with blueshifted (anti-Stokes) waves through a process known as coherent anti-Stokes Raman scattering (CARS). Under the proper conditions, a significant amount of anti-Stokes light is produced[1], which is useful for wavelength conversion[6].

The Raman effect can also be harnessed to produce optical loss through a process known as inverse Raman scattering (IRS)[7,8]. In this process, light injected at the anti-Stokes wavelength experiences attenuation in the presence of an intense pump. This resonant attenuation of the anti-Stokes wave increases with the intensity of the pump wave, opposite to the exponential gain of the Stokes wave during Raman amplification. IRS was first reported by Jones and Stoicheff in organic liquids[7], and has been used as a technique for Raman spectroscopy[9]. Since IRS occurs at the anti-Stokes wavelength, it avoids fluorescence contamination[7,9].

Apart from its utility in spectroscopic measurement, IRS could prove to be a valuable tool for photonic signal processing. However, to the best of our knowledge, IRS has never been observed in a semiconductor medium. In this letter, we demonstrate the resonant attenuation of an optical signal in a silicon waveguide through inverse Raman scattering. In agreement with the known Raman characteristics of silicon[10], the observed attenuation bandwidth is ~100 GHz, but in contrast to SRS, it is *blueshifted* by 15.6 THz from the pump wavelength.

IRS brings another important tool to the silicon photonics toolbox. For applications, two salient features of IRS are its wavelength flexibility and native all-optical platform. For example, chip-scale IRS may have important applications in photonic processing of high-bandwidth radio-frequency signals. To be insensitive to optical phase fluctuations, and hence to attain stable operation in the face of environmental variations, photonic signal processors operate on the optical power, as opposed to the electric field[11,12]. Unfortunately, it is difficult to perform key mathematical operations and processing functions such as subtraction of two intensity modulated signals because power is fundamentally a positive quantity[11,12]. IRS breaks this limitation: if the Raman pump is intensity modulated at radio frequencies, the anti-Stokes output will have the inverted modulation. In signal processing terminology, IRS provides a means to produce negative intensity taps, which are necessary for phase-insensitive subtraction of modulated signals. With a relatively fast response time of around 3 ps in silicon, IRS is a suitable candidate for performing all-optical switching and modulation functions. As a fast switch, silicon-based IRS may also find application in wavelength-division multiplexing (WDM) systems, which have been proposed to increase data throughput in optical interconnects.

In spontaneous Raman scattering, a minute amount of Stokes light and an even smaller amount of anti-Stokes light are spontaneously radiated, with the proportion determined by the thermal occupation factor of the excited vibrational state. When Stokes or anti-Stokes input waves are added, however, the situation is dramatically different: an input Stokes wave is amplified through SRS, while an input anti-Stokes wave suffers attenuation through IRS (cf. Figure 1). It is worth stressing that SRS and IRS are corollary processes, with similar spectral characteristics. Amplification of the Stokes field comes at the expense of pump, whereas, photons are transferred to the pump during the attenuation of anti-Stokes field.

As described above, Raman scattering also couples the Stokes and anti-Stokes fields through CARS[1], a process in which the power transfer between Stokes and anti-Stokes waves depends on the phase mismatch between the pump ($k_P$), Stokes ($k_S$), and anti-Stokes ($k_A$) propagation constants: $\Delta k = 2k_P - k_S - k_A$. When a Stokes signal is present, the attenuation suffered by an anti-Stokes signal through IRS is, therefore, influenced by the transfer of energy from the Stokes signal through CARS. Even in the presence of phase mismatch, significant transfer of energy to the anti-Stokes wave can occur for high pump intensity[13]. This can be reduced by eliminating Stokes input light; however, at high pump power, four-wave mixing coupled with Raman amplification can still produce substantial Stokes radiation, affecting the anti-Stokes attenuation as discussed in more detail below.

In silicon, the landscape is complicated by additional nonlinear effects. In the presence of intense fields, two-photon absorption (TPA) produces broadband optical attenuation, and generates free carriers, which cause further free-carrier absorption (FCA)[14]; these effects result in a self-limiting process that depletes intensity of the pump wave[15]. Simultaneously, self-phase modulation (SPM), produced by concerted action of the Kerr nonlinearity and free-carrier refraction (FCR), shifts the pump spectrum towards blue wavelengths[15]. As the lifetime of free carries is ~1-10 ns in silicon waveguides, carrier build up may be minimized using short pulses with repetition period much longer than the carrier lifetime. Although FCA is detrimental to the operation of silicon Raman amplifiers, we demonstrate that it surprisingly facilitates the observation of IRS.

In our experiments, picosecond optical pulses from a mode-locked laser operating at 1550 nm are split to produce pump and probe pulses. One portion is stretched, amplified, and compressed to generate 20 ps pump pulses with 2 nm bandwidth and peak power up to 230 W.

The other portion is amplified and sent through a nonlinear fiber to generate a flat optical continuum centered at the anti-Stokes wavelength (~1433 nm). This continuum is filtered to a ~30 nm bandwidth to ensure that the input signal does not contain any energy at the Stokes wavelength. The pump and anti-Stokes pulses are coupled to a silicon waveguide, and the IRS spectrum is measured at the output of the waveguide.

Experimentally measured IRS spectra are shown in Figure 2A. At the highest pump intensity, the signal minimum becomes comparable to the noise floor of the spectrum analyzer. The broadband reduction in anti-Stokes power with increasing pump intensity arises from TPA (one pump photon and anti-Stokes photon), and FCA due to carriers liberated by the pump. The resonant IRS attenuation (discounting the broadband absorption) increases exponentially with pump power initially, as expected, but levels off and decreases slightly for high power (cf. Figure 2B). Nevertheless, IRS attenuation values >15 dB are readily observed.

At high pump intensities, we observe the generation of Stokes signal (cf. Figure 2A, inset). The generation of this signal arises primarily from Raman amplification of power transferred from the anti-Stokes probe to the Stokes wavelength by coherent four-wave mixing. This Stokes signal is also transferred back to the anti-Stokes wavelength via CARS, which tends to limit the observable IRS attenuation. This process coupled with the self-limiting nature of the pump power are likely responsible for the high-power reduction in IRS absorption illustrated in Figure 2B. These issues are discussed further in as the context of our numerical studies presented below. It may also be noted that the IRS spectrum shifts towards shorter wavelengths as the pump intensity is increased, which results from the SPM-induced blueshift of the pump signal as shown in Figure 3.

We perform numerical simulations of IRS in silicon using the generalized nonlinear Schrödinger equation (NLSE). This method has been successfully applied to the modeling of pulsed SRS in silicon waveguides[16,17,18]. We include the instantaneous electronic nonlinearity of the medium, the delayed vibrational (Raman) response, as well as TPA, FCA, and FCR. Free-carrier generation is solved simultaneously with the NLSE to determine the carrier concentration and the impact of the carriers on the electromagnetic field (see Appendix). We start with a narrowband pump pulse, add a synchronous weak broadband anti-Stokes probe pulse, and monitor the anti-Stokes attenuation, as in the experiment.

In Figure 4A, we show the calculated attenuation at the anti-Stokes wavelength vs. propagation distance within the waveguide. Here, we see that the attenuation increases exponentially early in the waveguide, but begins to level off with propagation distance. As the pump power is increased, the attenuation develops oscillatory structure and reaches a limiting level more quickly. As described above, CARS contamination and the self-limiting nature of the pump limit the IRS attenuation. CARS transfer can limit the IRS attenuation because a significant Stokes signal builds up even when there is no input Stokes wave. Once created, coherent Raman scattering cyclically transfers energy back and forth between the Stokes and anti-Stokes wavelengths on a length scale that depends on the phase mismatch: $L = 2\pi / \Delta k$. Since the phase mismatch here is dominated by material dispersion, we have $L = 2\pi / \beta_2 \omega_v^2$, where $\beta_2$ is the group-velocity dispersion and $\omega_v$ is the Raman frequency shift. Given silicon's material dispersion, $\beta_2 = 1.2 \, \text{ps}^2/\text{m}$, we find an oscillation length of $L = 0.54 \, \text{mm}$, which matches the period observed in Figure 4A. Figure 4B shows the anti-Stokes attenuation at the waveguide output as a function of the input pump power. As also seen in the experiment, the attenuation increases to a maximum, but begins to decline at high pump power.

It is worth noting that the IRS attenuation is generally subject to limitation. If the dispersion is large, the anti-Stokes and Stokes waves are decoupled, which inhibits the CARS contamination of the IRS absorption. On the other hand, larger dispersion also causes the pump and anti-Stokes fields to walk off more rapidly, limiting the IRS interaction by a different means. For smaller dispersion, the phase mismatch is smaller, and CARS contamination takes hold more rapidly. Attempting to increase the pump power enhances the contamination process, and also fuels the power-limiting nonlinear absorption of the pump itself. Furthermore, as the anti-Stokes signal is attenuated through IRS, it becomes increasingly vulnerable to CARS contamination because the Stokes wave grows while the anti-Stokes level falls; the transfer of a small fraction of the growing Stokes wave has a substantial impact on the depleted anti-Stokes signal.

Interestingly, we find that free-carrier effects actually facilitate the observation of IRS under the present circumstances. Without FCA, the pump power remains larger throughout the waveguide, favoring the growth of the Stokes field and increasing the CARS conversion efficiency. For relatively moderate power levels, CARS contamination becomes so strong that resonant gain is observed at the anti-Stokes wavelength, overwhelming the IRS induced attenuation. As a result of pump spectral modifications, the CARS and IRS spectra do not overlap perfectly, leaving a small amount of residual attenuation (cf. Figure 5); however, experimentally, this attenuation would be rather difficult to identify apriori as the gain becomes the dominant spectral feature.

In summary, we have observed inverse Raman scattering in a semiconductor medium for the first time. IRS causes the attenuation of anti-Stokes energy when it propagates with an intense optical pump wave in silicon. Attenuation values in excess of 15 dB have been observed with pump intensities of the order of 4 GW/cm$^2$. Inverse Raman scattering may be very useful for optical signal processing on silicon chips, and is an important complimentary tool to Raman gain and wavelength conversion via SRS and CARS.

**Acknowledgements**

**Appendix**

In our experiments, the pump and anti-Stokes pulses are combined in free space and coupled to a silicon waveguide with mode area of ~2.8 μm², through a microscope objective. The input coupling loss of the setup, including reflection and mode-mismatch loss, is estimated to be approximately 6.6 dB. The 2 cm long optical waveguide has a linear propagation loss of ~0.5 dB/cm. An optical delay line in the path of the anti-Stokes signal is used to adjust the relative delay between the pump and probe pulses, and is critical for optimizing the efficiency of the attenuation observed through IRS. At the output of the waveguide, light is collected using a single-mode fiber and fed to an optical spectrum analyzer.

Our numerical simulations utilize the generalized nonlinear Schrödinger equation for the field envelope:

$$\frac{\partial A(z,t)}{\partial z} + \frac{i\beta_2}{2}\frac{\partial^2 A(z,t)}{\partial t^2} = i\gamma\left[1+\frac{i}{\omega_0}\frac{\partial}{\partial t}\right]A(z,t)\int_{-\infty}^{t} R(t-t')|A(z,t')|dt' + \left(\frac{i\omega_0 n_{FCR}}{c} - \frac{1}{2}\alpha_{FCA}\right)A(z,t),$$

where $\beta_2 = 9.5\,\text{fs/nm/cm}$ describes the dispersion of bulk silicon, $\gamma$ is the nonlinear coefficient, $R(t)$ is the third-order nonlinear response function, $\alpha_{FCA}$ is the free carrier absorption coefficient, and $n_{FCR}$ is the free carrier contribution to the refractive index[16-18]. The nonlinear coefficient is given by: $\gamma = \omega_0 n_2 / cA_{eff} + i\beta_{TPA}/2A_{eff}$, where $n_2 = 6\times 10^{-5}\,\text{cm/GW}$, $\beta_{TPA} = 0.5\,\text{cm/GW}$, and $R(t) = (1-f_R)\delta(t) + f_R h(t)$. The response function $R(t)$ includes electronic (instantaneous) and vibrational (delayed) nonlinearities, and the weighting factor $f_R$ is calculated to normalize the time-integrated response. In the Fourier domain, the delayed portion of the response can be related to the Raman gain function: $\text{Im}[\tilde{H}(\Omega)] = \dfrac{g(\Omega)}{2k_0 n_2 f_R}$, where $\Omega$ is the frequency relative to the pump $\omega_P$. The gain has frequency dependence: $g(\Omega) \propto \dfrac{\omega}{n(\omega)} \mathrm{Im}\, \chi_R(\Omega)$, where $\omega = \Omega + \omega_P$,

$n(\omega)$ is the refractive index, and $\chi_R(\Omega) = \dfrac{2\Omega_S \Delta\omega}{\Omega_S^2 - \Omega^2 + 2i\Omega\Delta\omega}$ is the normalized Raman susceptibility. The Raman frequency shift and linewidth are $\Omega_S$ and $2\Delta\omega$, respectively. Since the Raman gain coefficient $g_R$ is typically known at the Stokes wavelength, we normalize the gain function to that value at the proper frequency. Under the approximation $n(\omega_S) \approx n(\omega_{aS})$, this produces $\tilde{H}(\Omega) = \dfrac{g_R}{2k_0 n_2 f_R} \dfrac{(\Omega + \omega_P)}{(-\Omega_S + \omega_P)} \chi_R(\Omega)$. We use a Raman gain coefficient of

$g_R = 7$ cm/GW at the Stokes wavelength, which is conservatively within the range of published values[4,6], a Raman linewidth of 105 GHz, and a Raman shift of 15.6 THz[10]. When quantifying the IRS absorption, we smooth the spectrum to reflect a finite spectral measurement resolution of ~1 nm.

In order to determine the free-carrier absorption and refraction, the concentrations of free electrons and holes are calculated using the independent differential equation that describes the generation of free carriers from TPA:

$\partial N_{e=h}(z,t)/dt = \beta_{TPA} |A(z,t)|^4 / 2\hbar\omega_0 A_{eff}$. We then calculate the free-carrier absorption coefficient $\alpha_{FCA}$ and refractive index change $n_{FCR}$ using the following empirical formulae[19,20]:

$$n_{FCR} = -\left(8.8 \times 10^{-22} N_e + 8.5 \times 10^{-18} N_h^{0.8}\right)$$

$$\alpha_{FCA} = 8.5 \times 10^{-18} N_e + 6.0 \times 10^{-18} N_h,$$

where $N_e$ and $N_h$ are the densities of electrons and holes, respectively in units of cm$^{-3}$.

**Figure 1**

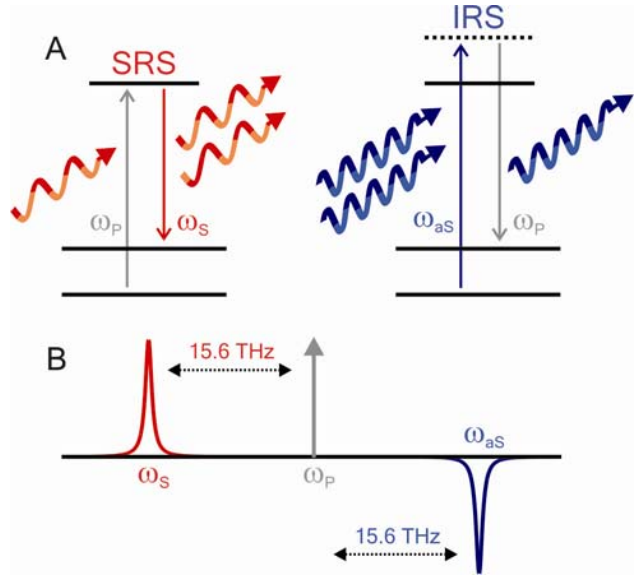

**Fig. 1. Schematic of inverse Raman scattering (IRS) and stimulated Raman scattering (SRS).** IRS and SRS are corollary processes arising in Raman scattering. A) In SRS, photons at the Stokes frequency $\omega_S$ are amplified at the expense of the pump at frequency $\omega_P$. In IRS, photons at the anti-Stokes frequency are *absorbed* in the *presence* of the pump. B) Raman scattering is a resonant process: the redshifted SRS gain line and the blueshifted IRS absorption line straddle the pump frequency, spaced by 15.6 THz in silicon.

**Figure 2**

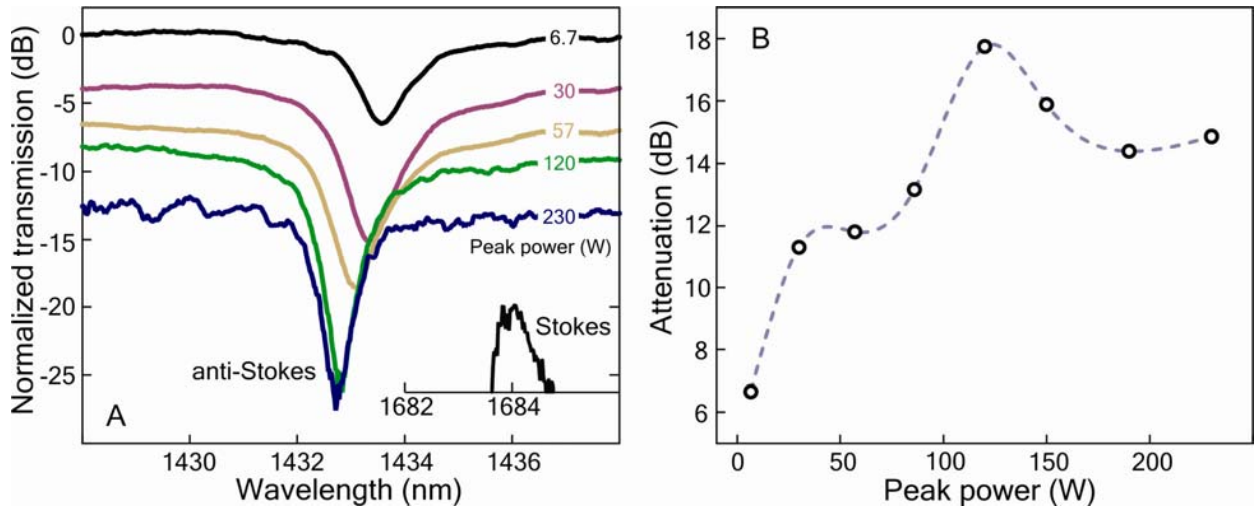

**Fig. 2. Experimental observation of inverse Raman scattering (IRS) in silicon waveguides.**
A) Normalized transmission of light vs. wavelength at the indicated input pump power levels (pump: $\tau = 20$ ps, $\Delta\lambda = 2$ nm). The anti-Stokes signal shows resonant attenuation, whose strength depends on the pump power—the hallmark of IRS. There is also power-dependent broadband loss arising from the two-photon and free-carrier absorption processes. The inset shows the transmitted Stokes signal (same scale). As there is no Stokes input, this signal arises from Raman amplification of both spontaneous emission and power transferred coherently from the anti-Stokes wavelength. B) Maximum *resonant* attenuation of the anti-Stokes signal vs. peak power. At high peak power, the resonant attenuation begins to decrease, as explained in the text.

**Figure 3**

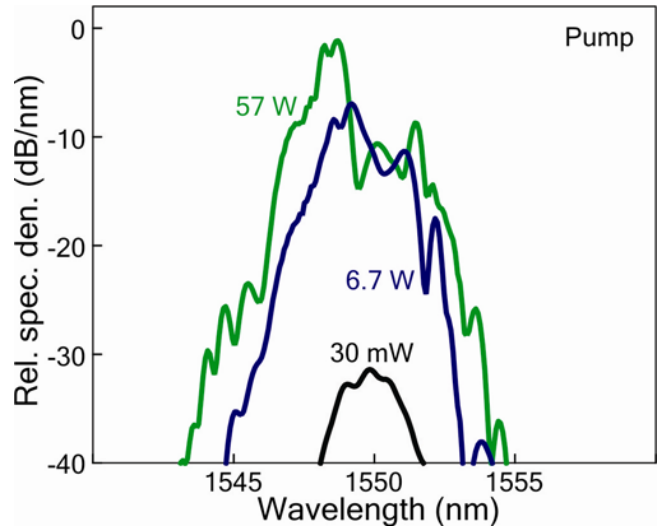

**Fig. 3. Measured spectrum of the pump light at the waveguide output.** Self-phase modulation, which is produced by both the Kerr nonlinearity and free-carrier refraction in silicon, modifies the output pump spectrum. The combined nonlinear modification blueshifts the spectral peak as the input power is increased, which causes the anti-Stokes absorption line to shift to shorter wavelengths as seen in Figure 2.

**Figure 4**

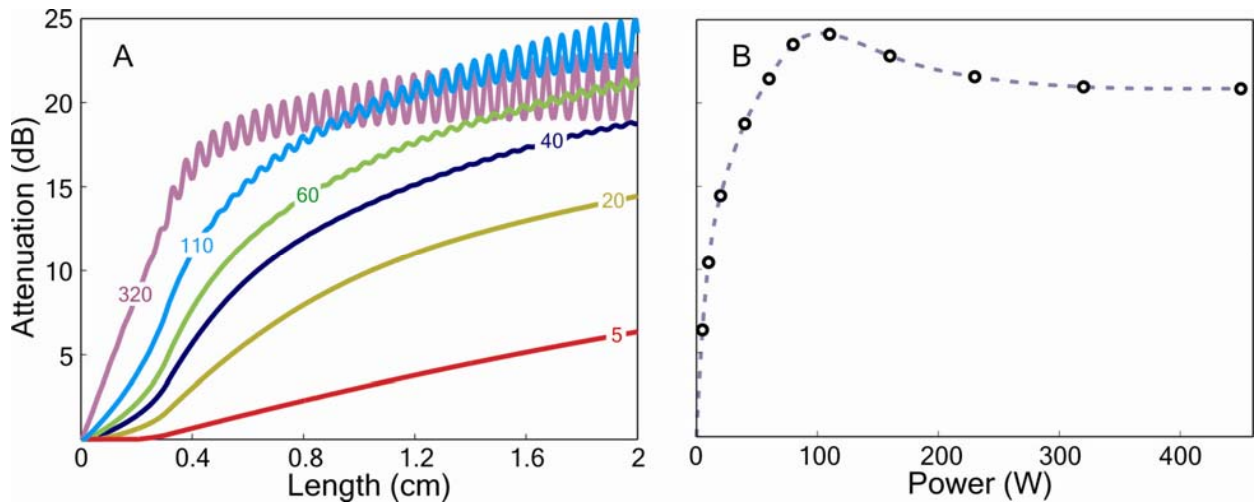

**Fig. 4. Simulation of inverse Raman scattering in a silicon waveguide.** A) Resonant attenuation vs. propagation length at the indicated pump power levels. At higher pump power levels, the attenuation rises rapidly near the start, but levels off further into the waveguide and develops an oscillatory structure due to coherent power transfer. B) Resonant attenuation at the waveguide output vs. input pump power. The attenuation rises rapidly at low pump power, but reaches a maximum and settles on a lower value at high power.

**Figure 5**

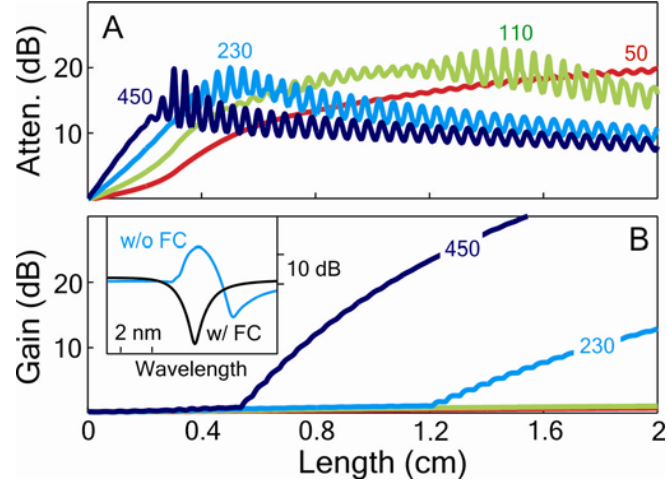

**Fig. 5. Simulation of the anti-Stokes signal in the absence of free-carrier absorption.** A) Resonant attenuation from inverse Raman scattering (IRS) and B) gain from coherent anti-Stokes Raman scattering (CARS) vs. propagation length at the indicated pump power levels. Due to pump spectral modifications and dispersion, the CARS and IRS spectra do not overlap perfectly, and anti-Stokes gain and attenuation can occur simultaneously. In the absence of free carriers, the pump power must be relatively small to observe significant IRS at the end of the waveguide. For higher power, the attenuation drops substantially, while the CARS gain increases, becoming the dominant spectral feature. The inset illustrates anti-Stokes spectra at 230 W with and without free-carrier effects.